\documentclass[aps,prb,twocolumn,superscriptaddress,floatfix]{revtex4-1}
\usepackage{graphicx}
\usepackage{amssymb}
\usepackage{amsmath}
\usepackage{upgreek}
\usepackage{dsfont}
\newcommand{\vect}[1]{\boldsymbol{\mathbf{#1}}}       
\newcommand{\mathcomma}{~ ,}            
\newcommand{\mathperiod}{~ .}           
\newcommand{\dosf}{\rho_{\mathrm{F}}}     
\def\bsigma{{\boldsymbol \sigma}}
\DeclareMathOperator{\sgn}{sgn}
\DeclareMathOperator{\tr}{tr}
\begin{document}
\title{Effect of weak disorder on the phase competition in iron pnictides}
\author{M.\ Hoyer}
\affiliation{Institut f\"ur Theorie der Kondensierten
Materie, Karlsruher Institut f\"ur Technologie, D-76131 Karlsruhe, Germany}
\author{S.\,V.\ Syzranov}
\affiliation{Institut f\"ur Theorie der Kondensierten
Materie, Karlsruher Institut f\"ur Technologie, D-76131 Karlsruhe, Germany}
\affiliation{Department of Physics, University of Colorado, Boulder, CO 80309, USA}
\author{J.\ Schmalian}
\affiliation{Institut f\"ur Theorie der Kondensierten
Materie, Karlsruher Institut f\"ur Technologie, D-76131 Karlsruhe, Germany}
\affiliation{Institut f\"ur Festk\"orperphysik, Karlsruher Institut f\"ur Technologie, D-76021 Karlsruhe, Germany}
\date{\today}
\begin{abstract}
We analyze the effect of weak disorder on the competition between antiferromagnetic order and superconductivity in a model for iron-based
superconductors. Under the assumption of an approximate particle-hole symmetry we show that conventional $s^{++}$~superconductivity cannot be realized in the case of coexisting magnetic and superconductive orders, observed experimentally at intermediate doping levels. 
This result holds for arbitrary impurity concentrations, and, in particular, in the clean limit. 
The inclusion of disorder further amplifies the phase competition between itinerant antiferromagnetism and conventional superconductivity. In addition, we analyze the effect of disorder on the characteristic length scales of the two order parameters, and find that in a disordered sample the staggered moment fluctuates on shorter scales than the superconductive order parameter, even if both length scales are the same in the clean limit. 
\end{abstract}
\maketitle
\section{Introduction}
Phase competition is a hallmark of strongly correlated electron systems
that exhibit ground states with rather distinct order yet of comparable
energy. Changing parameters in the Hamiltonian by applying external
fields, stress, or chemical composition allows one to tune from one state to another.
Complex phase diagrams divulge, on the one hand,
our limitation to make quantitative predictions for a given compound.
On the other hand, the nature of the competing ordered states reflects
the relevant degrees of freedom in a low energy description. 
An important aspect of phase competition that is crucial for a realistic description of correlated materials is the role of disorder and impurities. 

\begin{figure}[b]
  \includegraphics[width=\linewidth]{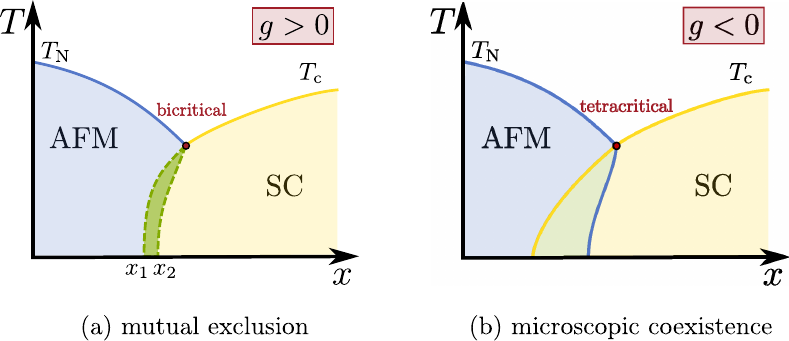}
 \caption{(Color online) Sketches of the two types of phase diagrams experimentally observed in iron-based superconductors. 
 (a)~Phase diagram where SC and AFM orders mutually exclude each other. The transition between AFM and SC is first-order, and there may be a region of 
 heterogeneous coexistence, depending on the thermodynamic variable we used as a control parameter.
 (b)~Phase diagram exhibiting a region where SC and AFM orders coexist microscopically, and thus compete for the same electrons. SC and AFM transitions are second-order, and meet in a tetracritical point.}
  \label{fig:phasediagrams}
\end{figure}
Iron-based superconductors display a phase diagram characterized
by antiferromagnetism, nematic order, and superconductivity along
with regions in the parameter space where the non-magnetic normal state
displays quantum critical and more conventional behavior\cite{Johnston2010,Paglione.Greene.2010,PhysicaC.469.614,Fernandes2014-Review}. Numerous
arguments support a sign-changing superconducting state, where the
$s^{+-}$~state, with opposite sign of the Cooper pair wave function
on hole and electron pockets, is the most prominent example\cite{HirschfeldMazin2011,Chubukov2012-Review}. Among the strongest
evidence in favor of this state are the emergence of a spin-resonance
mode in inelastic neutron scattering experiments\cite{Christianson2008-SpinResonance,InosovBourgesKeimer2009-SpinResonance} and the field-dependence
of the quasi-particle interference pattern in scanning tunneling spectroscopy\cite{Hanaguri2010}.
In Refs.~\onlinecite{PhysRevB.81.140501,PhysRevB.81.174538,PhysRevB.82.014521} it was argued that the nature of the phase competition
between antiferromagnetism and superconductivity can be another powerful
tool to distinguish between sign-changing and sign-preserving superconducting states. This
conclusion was based on two key ingredients: i) the same electrons
that contribute to the ordered antiferromagnetic moment also contribute to
the Cooper pair condensate and ii) there exists at least an approximate
particle-hole symmetry between the electron and hole bands shifted
by the ordering vector of the magnetic order $\mathbf{Q}$:
\begin{equation}
\xi_{\mathrm{hole}}(\mathbf{k})\approx-\xi_{\mathrm{elec.}}(\mathbf{k}+\mathbf{Q}).
\end{equation}
This latter condition seems to be reasonably well satisfied in many iron-based superconductors, allowing one to make the connection between phase stability and pairing state, a conclusion that cannot be drawn so easily in other systems. 
If both conditions are fulfilled, Refs.~\onlinecite{PhysRevB.81.140501,PhysRevB.81.174538,PhysRevB.82.014521} concluded that antiferromagnetism
and conventional ($s^{++}$) superconductivity will be separated
by a strong first-order transition as sketched in Fig.~1a. AFM and SC phase transition lines meet at a bicritical point. 
In contrast, in the case of $s^{+-}$~pairing (and
similarly for $d$-wave pairing), the system is at the verge between
a first-order transition and crossing second-order transitions with
a regime of homogeneous and simultaneous order of both states as depicted in Fig.~1b. 
Depending on details of the electronic structure and interactions, 
AFM and SC phase transition lines meet at a bicritical point or at a tetracritical point, respectively. The
observation of crossing second-order lines in some, but not all, systems
was then argued to be strong evidence for unconventional pairing.
Thus, one can ``read off'' the superconducting pairing state from
the phase diagram of the iron-based superconductors. 
At the heart
of this conclusion was the fact that the magnetic order parameter
has a typical momentum $\mathbf{Q}$ which couples the superconducting
condensates in electron and hole sheets of the Fermi surface, similar
to an internal Josephson coupling in momentum space. The effect is
rooted in the same coherence factors that lead to the resonance mode
enhancement at the transferred momentum $\mathbf{Q}$ in inelastic 
neutron scattering experiments. 

The theory of Refs.~\onlinecite{PhysRevB.81.140501,PhysRevB.81.174538,PhysRevB.82.014521} was based on the assumption that disorder
plays no role in the phase competition. However, the widely studied
system Ba(Fe$_{1-x}$Co$_{x}$)$_{2}$As$_{2}$ and a number of related
systems are clearly affected by disorder. This insight was revealed
in first-principle calculations\cite{WadatiSawatzky2010,PhysRevLett.108.207003}, and very clearly demonstrated in recent
NEXAFS experiments that identified the spin and valence state of Co\cite{MerzSchupplerLoehneysen2012}.
A rigid band calculation of a perfectly clean system as performed in Refs.~\onlinecite{PhysRevB.81.140501,PhysRevB.81.174538,PhysRevB.82.014521} is therefore not sufficient. Disorder has been shown to strongly affect
the phase competition between antiferromagnetism and superconductivity. 
It is therefore crucial to investigate the role of disorder on the
interplay between superconducting pairing and phase competition. 
More generally, the investigation of the role of disorder in unconventional superconductors has proven to yield important clues with regard to the competition between alternative states of order\cite{KohnoSigrist1999,ZhangSachdev2002,Balatsky2006,Galitski2008,Kogan2009,Alloul2009,Hardy2010,PhysRevB.84.214521,MoorEfetov2011,FernandesChubukov2012}. 
In addition, the more general question of how one can manipulate the
degree of competition between different phases clearly deserves more
detailed attention.

In this paper, we investigate the role of disorder on the phase competition
between magnetism and superconductivity. 
In our analysis, we therefore consider the regime where the ordering temperatures of both states remain finite, i.e., are not suppressed due to disorder. 
By analyzing several experimentally
motivated models for the microscopic nature of disorder we demonstrate
that the distinct phase competition between itinerant antiferromagnetism
and either $s^{++}$ or $s^{+-}$ superconductivity is further enhanced
if one includes disorder. 
Thus, the statements of Refs.~\onlinecite{PhysRevB.81.140501,PhysRevB.81.174538,PhysRevB.82.014521}, relating
the phase diagram and the nature of the pairing state,
apply as well to disordered systems.
\section{Phenomenological approach}\label{sec:phenomenology}
Before we enter a microscopic analysis of the phase competition, we phenomenologically describe the situation of two second order phase transitions, one antiferromagnetic and one superconductive, that meet in a multicritical point.
The resulting phase diagrams can be divided in two classes: Either superconductivity is able to microscopically coexist with antiferromagnetism, or superconductivity and antiferromagnetism mutually exclude each other. In Fig.~\ref{fig:phasediagrams}, these two types of phase diagrams found in iron pnictide materials are sketched.
The first class shows a bicritical point in the phase diagram, and a first-order transition line between superconductivity and antiferromagnetism, Fig.~1a. 
The second class shows a tetracritical point in the phase diagram, a region where both order parameters are nonzero (referred to as \emph{homogeneous coexistence}~\cite{PhysRevB.81.140501,PhysRevB.81.174538,PhysRevB.82.014521}), and all phase-transitions are second-order, Fig.~1b. 
 
Around the multicritical point in the phase diagram, the free energy can be expanded simultaneously in both order parameters, in the spirit of Ginzburg-Landau theory of superconductivity~\cite{GinzburgLandau1950,Gorkov1959}.
The most generic form of the free energy, allowed by the symmetry, in terms of
the antiferromagnetic ${\bf M}$ and the superconductive $\Delta$ order parameters reads as
\begin{align}
 \Delta F &= \int\mathrm{d}\vect{r}\,\bigg[ \tfrac{1}{2}a_\mathrm{m}\vect{M}^2+\tfrac{1}{4}u_\mathrm{m}\vect{M}^4 \nonumber \\  
  & \quad  +\tfrac{1}{2}a_{\mathrm{s}}|\Delta|^2  +\tfrac{1}{4} u_{\mathrm{s}}|\Delta|^4+\tfrac{1}{2}\gamma\vect{M}^2|\Delta|^2
  +\ldots
  \bigg].
 \label{eq:free-energy-expansion}
\end{align}
The last quartic term in Eq.~(\ref{eq:free-energy-expansion}) has to be positive, $\gamma>0$, in order
to ensure competition between the two ordered phases.
Gradient terms accounting for temporal and spatial fluctuations of the order parameters can be included in this expansion of the free energy as well, and will be discussed later. 
Whether the phase transitions are first or second order, and thereby the shape of the phase diagram near the multicritical point, is determined by the quartic coefficients. 
The analysis of the quadratic form associated with the quartic terms suggests to introduce the quantity 
\begin{equation}
  g=\frac{\gamma}{\sqrt{u_\mathrm{m}u_\mathrm{s}}}-1\mathcomma
\end{equation}
which determines the shape of the phase diagram. 
For $g<0$ we encounter a phase diagram with second-order phase transitions only, and thus a tetracritical point. 
For $g>0$, antiferromagnetism and superconductivity are separated by a first-order phase transition line that terminates in a bicritical point. 
Below, we will determine microscopic expressions for the coefficients in Eq.~\eqref{eq:free-energy-expansion} for a disordered system with electron--electron interactions. The resulting values of these coefficients determine the location of the multicritical point as a function of microscopic parameters. Changing the disorder strength or other parameters in the Hamiltonian will then affect the location of the multicritical point. Eq.~\eqref{eq:free-energy-expansion} is then valid in the vicinity of this multicritical point. 

In Ref.~\onlinecite{PhysRevB.81.140501,PhysRevB.81.174538,PhysRevB.82.014521}, using a weak coupling analysis, it was found for the clean system that $g_{++}=2$ and $g_{+-}=0$. 
In case of perfect particle-hole symmetry, the latter result is exact, as shown in Ref.~\onlinecite{EPL.88.17004}. 
In what follows, we derive the coefficients of the expansion~\eqref{eq:free-energy-expansion} from a microscopic model describing the essential features of the iron pnictides in presence of disorder. The model is introduced in section~\ref{sec:model} and its implications in the presence of disorder are studied in the remainder of this paper. 
\section{Model}\label{sec:model}
\begin{figure}
  \includegraphics[width=0.35\textwidth]{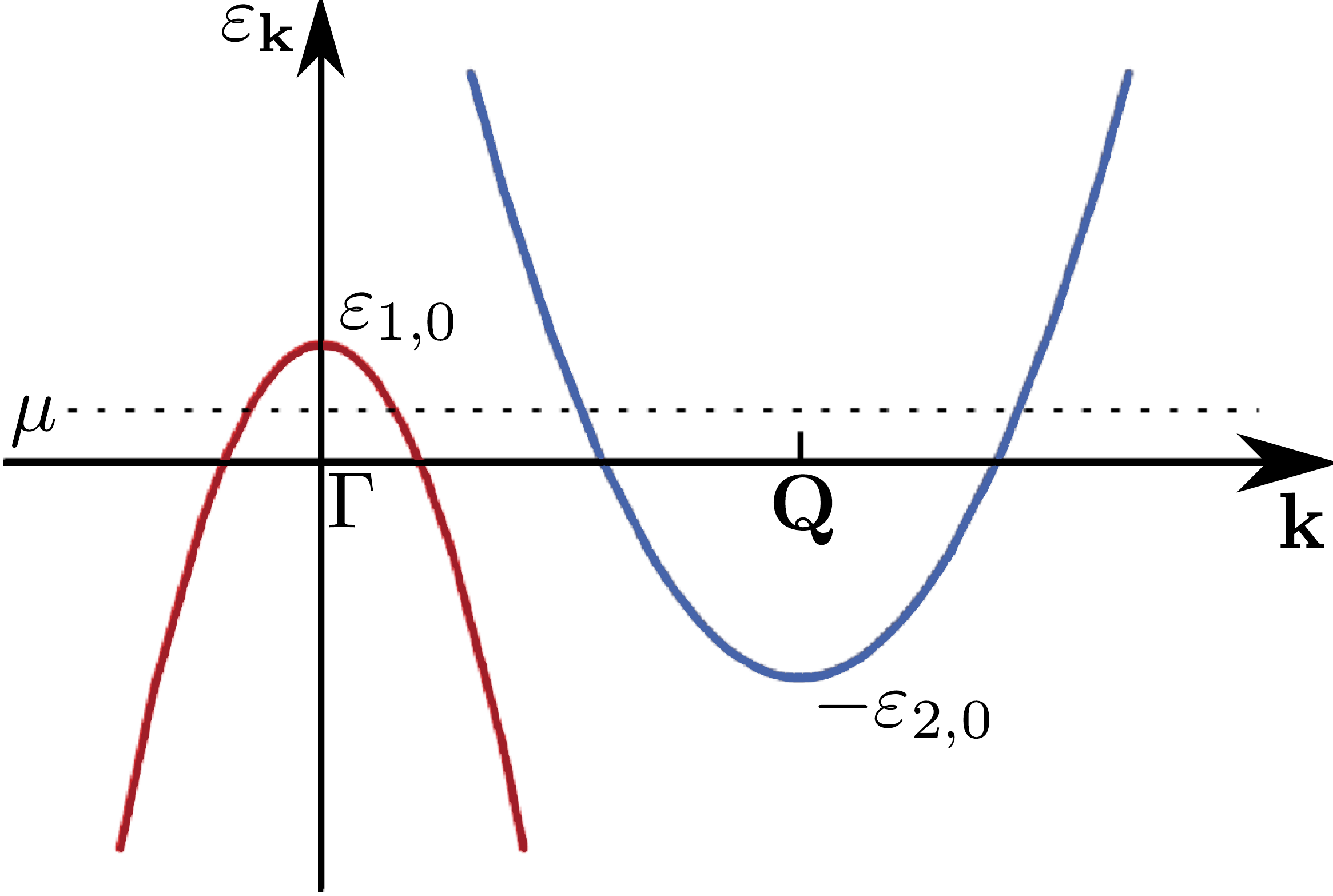}
  \caption{(Color online) The two-band model on which our microscopic description of the iron pnictides is based. The hole band at the $\Gamma$~point is a circular band whereas the electron band centered around $\vect{Q}$ could in principle be of elliptic form. Additionally we could introduce a finite chemical potential $\mu$ to fit our model to more realistic band structures, and introduce $\xi_{\alpha,\vect{k}}=\varepsilon_{\alpha,\vect{k}}-\mu$. }
  \label{fig:two-band-model}
\end{figure}
We consider the two-band model that is illustrated in Fig.~\ref{fig:two-band-model}. It consists of a circular hole band at the $\Gamma$ point, and an elliptical electron band shifted by vector $\vect{Q}$. This is the minimal model in which phase competition of antiferromagnetism and $s^{++}$ or $s^{+-}$ superconductivity in the iron pnictides can be studied. 
The Hamiltonian 
\begin{equation}
  \mathcal{H}=\mathcal{H}_0+\mathcal{H}_\mathrm{SC}+\mathcal{H}_\mathrm{AFM}+\mathcal{H}_\mathrm{dis}
  \label{eq:hamiltonian}
\end{equation}
of this two-band model contains the usual non-interacting part 
\begin{equation}
  \mathcal{H}_0=\sum_{\mathbf{k},\sigma}\sum_\alpha\xi_{\alpha,\mathbf{k}}\psi_{\alpha,\mathbf{k},\sigma}^\dagger \psi_{\alpha,\mathbf{k},\sigma}\mathcomma
\end{equation}
where we label the two bands by the index~$\alpha$. 
$\psi^\dagger_\alpha$ and $\psi_\alpha$ are the creation and annihilation operators in the respective band. 
Since small changes in the band structure lead to small changes of $g$, we focus on the particle-hole symmetric case in the following. 
Deviations from particle-hole symmetry were investigated in Refs.~\onlinecite{PhysRevB.81.140501,PhysRevB.81.174538,PhysRevB.82.014521}. Within the assumption of particle-hole symmetry, no further details of the dispersion $\xi_{\vect{k}}$ are needed to calculate $g$. In our analysis of the coefficients of the momentum dependence we assume  a parabolic dispersion, i.\,e., $\xi_{1,\vect{k}}=\xi_0-\tfrac{\vect{k}^2}{2m}$, and $\xi_{2,\vect{k}}=-\xi_{1,\vect{k}}$  in case of particle-hole symmetry. Note that band~2 is centered around $\vect{Q}$ which in our notation is included in the band index. 
\subsection{Superconductivity and magnetic order}
The electrons are subject to an effective electron--electron interaction leading to superconductivity and antiferromagnetism.
Here, we do not attempt to find microscopic expressions for the pairing interaction but are investigating the consequences of alternative pairing states.
So, we assume for simplicity that the superconductive and antiferromagnetic couplings are described by two
microscopic Hamiltonians of different nature.

Superconductivity is described by a BCS-like Hamiltonian, 
\begin{align}  
  \mathcal{H}_\mathrm{SC}&=\sum_{\mathbf{k},\vect{k}^\prime, \vect{q}}\sum_\alpha V^\mathrm{s}_{\vect{k},\vect{k}^\prime,\vect{q}} \psi_{\alpha,\mathbf{k}+\mathbf{q},\uparrow}^\dagger \psi_{\alpha,-\mathbf{k},\downarrow}^\dagger \nonumber \\ 
  &\quad \times \psi_{\overline{\alpha},-\mathbf{k}^\prime+\mathbf{q},\downarrow} \psi_{\overline{\alpha},\mathbf{k}^\prime,\uparrow} \mathcomma\label{eq:hamiltonian-sc}\\
 V^\mathrm{s}_{\vect{k},\vect{k}^\prime,\vect{q}} &= \left\{\begin{matrix} V_\mathrm{s} & \text{for } |\xi_{\vect{k}}|,|\xi_{\vect{k}^\prime}|,\xi_{\vect{k}+\vect{q}}|,|\xi_{-\vect{k}^\prime+\vect{q}}|<\Lambda_\mathrm{s} \mathcomma \\ 0 & \text{otherwise} \mathcomma\end{matrix}\right. \nonumber 
\end{align}
where $\bar{\alpha}$ refers to the opposite of $\alpha$. 
The electron--electron interaction leading to superconductivity is present for electrons with energies within a shell of width $2\Lambda_\mathrm{s}$ around the Fermi energy.
For phonon-mediated electron--electron interaction leading to conventional superconductivity, this energy cut-off would be given by the Debye frequency. In case of an electronic pairing mechanism, the cut-off is expected to be of the order of the Fermi energy. 
The same electrons that form the Cooper pairs are subject to an interaction that might lead to magnetic order. 
Antiferromagnetism shall be described in an itinerant picture, by 
\begin{align}
  \mathcal{H}_\mathrm{AFM}&=\sum_{\mathbf{k},\mathbf{k}^\prime,\mathbf{q}}\sum_{\sigma,\sigma^\prime,s,s^\prime}\sum_{\alpha,\beta} V^\mathrm{m}_{\vect{k},\vect{k}^\prime,\vect{q}} \psi_{\alpha,\mathbf{k},\sigma}^\dagger \psi_{\beta,\vect{k}^\prime,s}^\dagger\nonumber\\
  &\quad \times \bsigma_{\sigma\sigma^\prime}
  \bsigma_{ss^\prime}\psi_{\bar{\beta},\vect{k}^\prime-\vect{q},s^\prime}^{}\psi_{\bar{\alpha},\mathbf{k}+\mathbf{q},\sigma^\prime}^{}
  \mathcomma\label{eq:hamiltonian-afm} \\
  V^\mathrm{m}_{\vect{k},\vect{k}^\prime,\vect{q}} &= \left\{\begin{matrix} V_\mathrm{m} & \text{for } |\xi_{\vect{k}}|,|\xi_{\vect{k}^\prime}|,|\xi_{\vect{k}+\vect{q}}|,|\xi_{\vect{k}^\prime-\vect{q}}|<\Lambda_\mathrm{m} \mathcomma \\ 0 & \text{otherwise} \mathcomma \end{matrix}\right.\nonumber
\end{align}
where $\Lambda_\mathrm{m}$ is the characteristic energy cutoff for magnetic interactions.
Its nature is unimportant for our results, and in what follows we assume that it is of the order of or
smaller than the Fermi
energy, $\Lambda_\mathrm{m}\lesssim\xi_{k_\mathrm{F}}$.
\subsection{Disorder}
Weak quenched disorder is represented by the potential $U_{\alpha\beta}\left(\vect{r}\right) = \sum_{i=1}^{N_\mathrm{imp}} u_{\alpha\beta}\left(\vect{r}-\vect{R}_i\right)$ of $N_\mathrm{imp}$ randomly distributed identical impurities,    
\begin{align}
  \mathcal{H}_\mathrm{dis}&=\int\mathrm{d}\vect{r}\,\Psi_{\alpha,\sigma}^\dagger\left(\vect{r}\right)U_{\alpha\beta}\left(\vect{r}\right)\Psi_{\beta,\sigma}\left(\vect{r}\right)\mathcomma  
\end{align}
where $\Psi^{(\dagger)}(\vect{r})$ are the field operators in position space. 
Here $U_{\alpha\beta}\left(\vect{r}\right)$ is the impurity potential matrix element that may describe intraband scattering ($\alpha=\beta$) as well as interband scattering processes ($\alpha\neq\beta$). 
In the following we assume the impurity potential $u_{\alpha\beta}\left(\vect{r}-\vect{R}_i\right)$ to be short-ranged. 
We describe impurity scattering in the Born approximation, in which the rate of electron collisions with impurities is characterized by 
\begin{equation}
  \frac{1}{\tau_{\alpha\beta\gamma\delta,\vect{k}}}=2\pi \int\frac{\mathrm{d}\vect{k}^\prime}{(2\pi)^2}\,n_\mathrm{imp}u_{\alpha\beta,\vect{k}-\vect{k}^\prime} u_{\delta\gamma,\vect{k}-\vect{k}^\prime}^\ast \operatorname{\updelta}(\xi_{\vect{k}}-\xi_{\vect{k}^\prime})\mathcomma
    \label{eq:definition-tau}
\end{equation}
where $u_{\vect{k}-\vect{k}^\prime}$ is the Fourier component of the potential of a single impurity and $n_\mathrm{imp}$ the impurity concentration. 

We consider the model with a sufficiently smooth disorder, which leads to significantly smaller interband than intraband scattering. 
This hierarchy of scattering rates in iron pnictides is also supported by experiment\cite{PhysRevB.86.140505,PhysRevB.80.174512,PhysRevB.84.020513}. 
Without loss of generality we may assume $s$-wave scattering which corresponds to $\updelta$-correlated disorder when considering the scattering in one band. More general models of disorder will not change the results qualitatively, leading only to the replacement of the elastic scattering time by the transport scattering time. 
Then, the scattering amplitudes are characterized by constants $u_{11}$ ($u_{22}$) for scattering within band~1 (2), and by a constant $u_{12}$ for scattering between the bands. 
Impurity scattering in this two-band model is therefore characterized by intraband scattering rates ${\tau_1}^{-1}\equiv{\tau_{1111}}^{-1}$ and ${\tau_2}^{-1}\equiv{\tau_{2222}}^{-1}$, 
and the interband scattering rate ${\tau_{12}}^{-1}\equiv{\tau_{1221}}^{-1}$, which can be also assumed momentum-independent. 
From Eq.~\eqref{eq:definition-tau} follows that the scattering rate ${\tau_{1221}}^{-1}$ is real, whereas ${\tau_{1212}}^{-1}$ is allowed to have a nontrivial phase. Both scattering rates have equal magnitude, $|{\tau_{1212}}^{-1}|={\tau_{1221}}^{-1}$. 
Note, the generalization of this approach to more extended impurities, discussed in Refs.~\onlinecite{Gastiasoro2013-Dimers} and~\onlinecite{AllanDavis2013}, will be done elsewhere\cite{unpublished}. 

In this paper, we consider several models of disorder to describe the corresponding physically relevant limits. 

\emph{A. Dominant hole-band scattering.} 
Several experiments and first-principles calculations have demonstrated that intraband scattering in the hole band (labeled by~1) is significantly stronger than in the electron band~(band~2). 
This is supported for example by transport measurements~\cite{PhysRevLett.103.057001,Fang2009}, scanning tunneling microscopy\cite{Science.336.563}, as well as first-principles density functional theory calculations\cite{PhysRevLett.108.207003}. 
The physically relevant limit is therefore ${\tau_2}^{-1}\ll{\tau_1}^{-1}$ for the corresponding intraband scattering rates. 
However, in this paper, we consider a more general model where the values of the intraband scattering rates are arbitrary, ${\tau_1}^{-1}\neq{\tau_2}^{-1}$. 

Furthermore, the transition temperature $T_\mathrm{c}$ is suppressed with increasing impurity concentration. In case of $s^{+-}$~pairing, this is caused by interband scattering processes\cite{Hirschfeld1988,Golubov.Mazin.1997}. 
The suppression of the SC~transition temperature is an order of magnitude smaller than theoretical prediction based on the scattering rates 
obtained from transport experiments\cite{PhysRevB.86.140505,PhysRevB.80.174512,PhysRevB.84.020513}. 
Hence the interband scattering rate ${\tau_{12}}^{-1}$ that leads to suppression of $T_\mathrm{c}$ is smaller than the intraband scattering rate in the band that dominates the transport properties\cite{}. 
Thus, in iron pnictides ${\tau_{12}}^{-1}<{\tau_1}^{-1}$ seems a reasonable starting point. 

Since the intraband scattering rate~${\tau_1}^{-1}$ in the hole band is clearly the largest scattering rate in iron pnictide materials, we neglect interband scattering in the first model. Since without interband scattering, the intraband scattering rates in the two bands simply add up in the physical observables calculated in this paper, a finite intraband scattering rate in the electron band yields qualitatively similar results, and we consider our model in the limit, 
\begin{equation}
  {\tau_1}^{-1}\neq{\tau_2}^{-1} \quad \text{and} \quad {\tau_{12}}^{-1}=0 \mathcomma
  \label{eq:model-a}
\end{equation}
where the physically relevant limit to iron pnictides is ${\tau_2}^{-1}\ll{\tau_1}^{-1}$. This model, summarized in Eq.~\eqref{eq:model-a}, will be referred to as model~A throughout this paper. 

\emph{B. Investigation of interband scattering.} 
Although the interband scattering seems to be significantly weaker than the intraband scattering in the hole band, setting the interband scattering rate to zero is an oversimplification with respect to some aspects. The weak suppression of the SC~transition temperature is one example of the consequences of a finite interband scattering rate. 
Therefore, we analyze the influence of a finite interband scattering rate as well, and use the model 
\begin{equation}
  {\tau_1}^{-1}={\tau_2}^{-1}\equiv{\tau_0}^{-1} \quad \text{and}\quad {\tau_{12}}^{-1}\equiv t{\tau_0}^{-1} \mathcomma
  \label{eq:model-b}
\end{equation}
for the investigation of interband scattering on the phase competition in iron pnictides. This model of disorder in iron pnictides has already been considered by Ref.~\onlinecite{PhysRevB.84.214521} in a slightly different context. 
We argued, that the interband scattering rate is smaller than the largest intraband scattering rate ${\tau_1}^{-1}$, thus the range $0< t< 1$ is the limit interesting for the ratio of interband to intraband scattering rates in iron pnictides. The model itself however is not limited to this parameter range, 
and allows for the analysis of arbitrary ratios~$t$. We will refer to this model with a finite interband scattering rate, summarized in Eq.~\eqref{eq:model-b}, as model~B in the following. 
\section{Full Ginzburg-Landau expansion}\label{sec:ginzburg-landau-expansion}
In this section, we derive as an illustration the full Ginzburg-Landau expansion for model~A of disorder in iron pnictides, based on the dominant scattering mechanism. Thus we consider the interband scattering rate~${\tau_{12}}^{-1}$ to be zero, and finite intraband scattering rates ${\tau_1}^{-1}$ and ${\tau_2}^{-1}$ in bands~1 and 2, respectively. Experimental evidence suggests that the hole band is more severely affected by impurities than the electron band, so for the sake of clarity we concentrate on the limit ${\tau_2}^{-1}=0$ in the following derivation of the free energy. We also calculated the expansion of the free energy for the more general case of arbitrary intraband scattering rates ${\tau_1}^{-1}\neq{\tau_2}^{-1}>0$. The results will be presented at the end of this section. 

\begin{figure}
  \includegraphics[width=\linewidth]{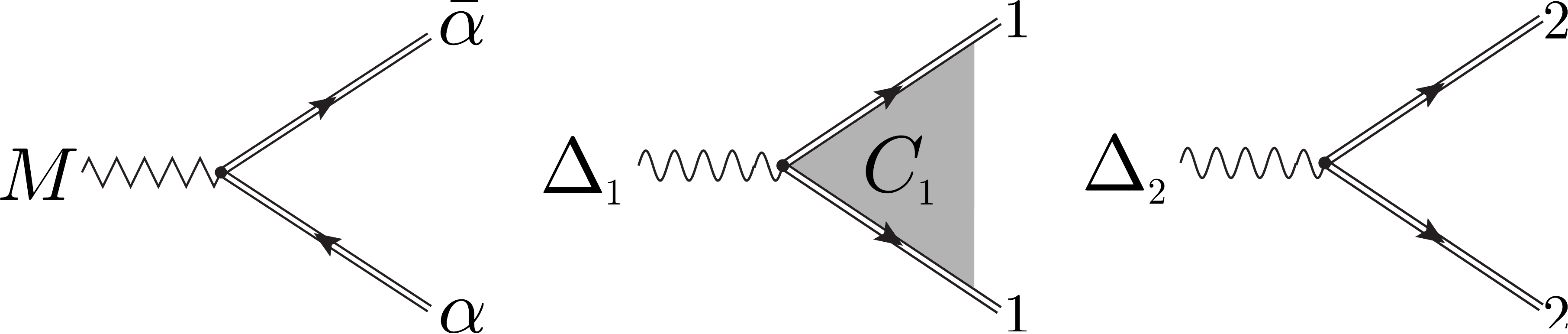}
  \caption{Vertices involved in the diagrams for the free energy.} 
  \label{fig:diagram-elem}
\end{figure}

One can calculate the coefficients in the Ginzburg-Landau expansion using the Eilenberger\cite{Eilenberger1968} approach or straightforwardly using perturbation theory based on the vertices depicted in Fig.~\ref{fig:diagram-elem}. The quadratic coefficients correspond to the diagrams presented in Fig.~\ref{fig:quadratic_coefficients}, and the quartic coefficients correspond to those shown in Figs.~\ref{fig:quartic-diagrams1} and~\ref{fig:quartic-diagrams2}. 

In order to arrive at an expansion of the free energy of the model~A introduced in section~\ref{sec:model}, we may write the partition function as 
\begin{equation} 
  \frac{\mathcal{Z}}{\mathcal{Z}_{\tilde{0}}}=\left< T_\tau\mathrm{e}^{-\int_0^\beta \mathrm{d}\tau\,\mathcal{H}_\mathrm{SC}\left[\psi^\dagger\left(\tau\right),\psi^{}\left(\tau\right)\right]+\mathcal{H}_\mathrm{AFM}\left[\psi^\dagger\left(\tau\right),\psi^{}\left(\tau\right)\right]} \right>_{\tilde{0}}
\end{equation}
in Matsubara interaction representation where the average ${\big<\ldots\big>}_{\tilde{0}} = \tr[\mathrm{e}^{-\beta\mathcal{H}_{\tilde{0}}}\ldots]/\mathcal{Z}_{\tilde{0}}$ refers to the non-interacting part $\mathcal{H}_{\tilde{0}}=\mathcal{H}_0 +\mathcal{H}_\mathrm{dis}$. 
We follow the usual procedure to make the action quadratic in the fermionic operators by introducing Hubbard-Stratonovich fields and find an effective action in terms of these fields. 
In momentum and frequency space, we use complex scalar fields $\Delta_{1,\vect{q}}\left(\omega_m\right)$ and $\Delta_{2,\vect{q}}\left(\omega_m\right)$ to decouple $\mathcal{H}_\mathrm{SC}$, and a three-component vector field $\vect{M}_{\vect{q}}\left(\omega_m\right)$ to decouple $\mathcal{H}_\mathrm{AFM}$, where $\omega_m$ are bosonic Matsubara frequencies. 
Thereby we find for the effective action in terms of these fields
\begin{align}
  S_\mathrm{SC}^\mathrm{eff}& =-\sum_{\vect{k},n}\sum_{\vect{q},m}\sum_{\alpha}\Big[\frac{1}{V_\mathrm{s}}\Delta_{\alpha,\vect{q}}^*(\omega_m)\Delta_{\overline{\alpha},\vect{q}}(\omega_m) \nonumber \\ 
  &\quad   + \Delta_{\alpha,\vect{q}}(\omega_m) \psi^\ast_{\alpha,\vect{k}+\vect{q},\uparrow}(\nu_n+\omega_m)\psi^\ast_{\alpha,-\vect{k},\downarrow}(-\nu_n) \nonumber \\ 
  &\quad + \Delta_{\alpha,\vect{q}}^*(\omega_m) \psi_{\alpha,\vect{k}+\vect{q},\downarrow}(\nu_n+\omega_m)\psi_{\alpha,-\vect{k},\uparrow}(-\nu_n)\Big]
  \label{eq:eff-hamiltonian-sc}
\end{align} 
and 
\begin{align}  S_\mathrm{AFM}^\mathrm{eff}&=-\sum_{\vect{k},n}\sum_{\vect{q},m}\Big[\frac{1}{V_\mathrm{m}}\vect{M}_{\vect{q}}^2(\omega_m)
+\sum_{s,s^\prime} \big\{ \vect{M}_{\vect{q}}(\omega_m)\vect{\sigma}_{ss^\prime} \nonumber \\ 
&\quad \sum_{\alpha} \psi_{\alpha,\vect{k},s}^\ast(\nu_n)\psi_{\overline{\alpha},\vect{k}+\vect{q},s^\prime}(\omega_m+\nu_n)\big\}\Big]\mathperiod
\label{eq:eff-hamiltonian-afm}
\end{align}
From the effective action we may construct the elements of a diagrammatic technique to derive the expansion of the free energy, and the superconducting and antiferromagnetic part of the action contain different types of vertices associated with the fields $\vect{M}$, $\Delta_\alpha$ and $\Delta_\alpha^*$, see Fig.~\ref{fig:diagram-elem}.   

In order to embed impurity scattering in the diagrammatic technique, we include impurity lines as new diagrammatic elements in our formalism, 
each associated with a factor $1/2\pi\dosf\tau_1$. 
Impurity scattering gives rise to a finite self energy and vertex corrections. 
Within this model of impurity scattering, the propagator in the electron band is given by the bare electron propagator, $G_{2,\vect{k}}(\nu_n)=(\mathrm{i}\nu_n-\xi_{2,\vect{k}})^{-1}$. The  propagator in the hole band is given by 
\begin{equation}
  \vcenter{\hbox{\includegraphics[height=0.75em]{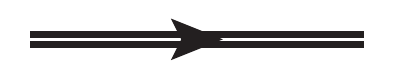}}}=G_{1,\vect{k}}(\nu_n) =\frac{1}{\mathrm{i}\nu_n-\xi_{1,\vect{k}}+\frac{\mathrm{i}}{2\tau_1}\sgn\nu_n} \mathcomma 
\end{equation}
with a finite self energy due to impurity scattering which we here treat in the lowest non-vanishing
order (Born approximation). 
Contributions with crossed impurity lines are neglected since they are suppressed by a small factor\cite{Abrikosov-Gorkov-Dzyaloshinski:MethodsOfQuantumFieldTheory} $1/k_\mathrm{F}l$ where $l=v_\mathrm{F}\tau_1$ 
is the mean free path, and $v_\mathrm{F}$ is the Fermi velocity. 
Therefore we construct our diagrams from vertices that already contain the full propagators. Additionally,  vertices are renormalized due to impurity scattering which is indicated by a shaded region in the diagrams.  
There are no vertex corrections of the vertices associated with~$\vect{M}$ and $\Delta_2$ to consider here since we set ${\tau_{12}}^{-1}=0$ and ${\tau_2}^{-1}=0$, respectively. 
Vertex corrections of vertices associated with $\Delta_1$ are given by the Cooperon ladder in band~1 which leads to a frequency-dependent factor 
\begin{align}
  C_1(\nu_n,\vect{q},\omega_m)&=\tfrac{2\tau_1|\nu_n|+1}{2\tau_1|\nu_n|} -\tfrac{1}{2}\tfrac{\tau_1^2v_\mathrm{F}^2}{(2\tau_1|\nu_n|+1)(2\tau_1|\nu_n|)^2}q^2 \nonumber \\  
  & \quad -\tfrac{\sgn\nu_n\tau_1}{(2\tau_1|\nu_n|)^2}\omega_m \mathcomma
\end{align}
to leading order in $\vect{q}$ and $\omega_m$. Here the restriction of $\sgn\left(\nu_n\right)\left(\nu_n+\omega_m\right)>0$ is implied in every summation that contains this Cooperon ladder. 
Again, in the calculation of the vertex corrections, and also in the construction of the diagrams, all contributions from crossed impurity lines can be neglected due to the small factor $1/k_\mathrm{F}l$. 
  
These prerequisites enable us to derive the full Ginzburg-Landau expansion of the free energy of our two-band model in presence of weak impurity scattering which reads 
\begin{align} 
  \Delta F&=\sum_{\alpha,\beta}\frac{a_{\mathrm{s},\alpha\beta}(\vect{q},\omega_m)}{2}\Delta_\alpha\Delta_\beta^*+\frac{a_\mathrm{m}(\vect{q},\omega_m)}{2}\vect{M}^2 \nonumber \\
  & \quad +\sum_{\alpha}\frac{u_{\mathrm{s},\alpha}}{4}\left|\Delta_\alpha\right|^4+\frac{u_\mathrm{m}}{4}\vect{M}^4 \nonumber\\ 
  & \quad +\sum_{\alpha\beta}\frac{\gamma_{\alpha,\beta}}{2}\vect{M}^2\Delta_\alpha\Delta_\beta^*
  \label{a1}
\end{align}
in frequency and momentum space. We note, that Eq.~(\ref{a1}) accounts for the  gradient terms $\propto(\nabla\Delta_\alpha)^2$, $\propto(\nabla\vect{M})^2$, $\propto(\partial_\tau\Delta_\alpha)$ and $\propto(\partial_\tau\vect{M})$ in the free energy, that characterize spatial and temporal fluctuations of the order parameters which is reflected in the dependence of the quadratic coefficients on finite incoming frequency~$\omega_m$ and momenta~$\vect{q}$. 

In our analysis, we implicitly assumed that the sole effect of disorder is to change the values of the coefficients of the order parameters. In the critical regime, it is well established that disorder may change the universality class of the transition, 
lead to Griffiths and quantum Griffiths effects, or even cause glassy behavior close to the transition point. 
These effects, however, only become important  in the very close vicinity of the critical point\cite{PhysRevLett.85.1532,Sknepnek2004} as a consequence of the weakness of the interaction sufficient to induce the ordered state. 
Thus, on the one hand, we consider our model sufficiently far away from the multicritical point to ignore these effects, while at the other hand sufficiently close to the multicritical point such that an expansion of the free energy is justified. 
Refs.~\onlinecite{PhysRevLett.85.1532,Sknepnek2004} demonstrated that this intermediate regime covers a wide range if the pairing and magnetic interactions are sufficiently weak. 
\subsection{Quadratic coefficients}
\begin{figure}
\includegraphics[width=\linewidth]{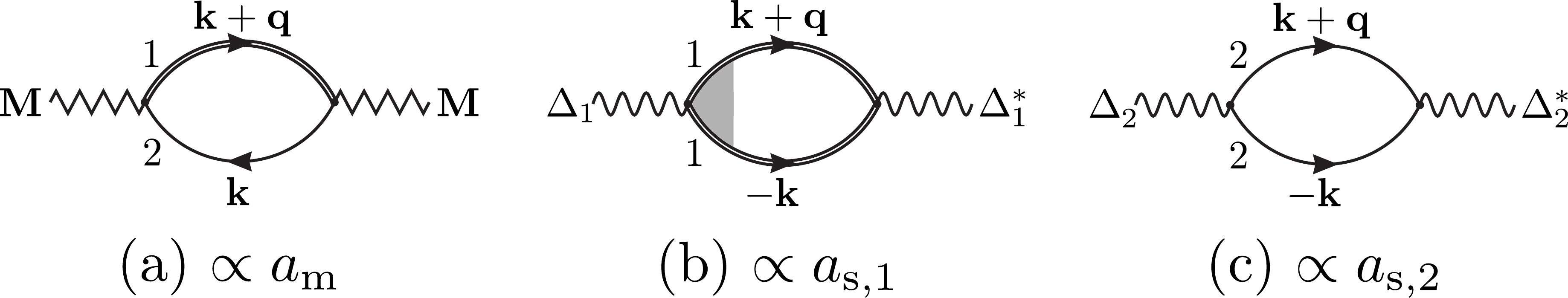}
  \caption{Diagrams for the quadratic coefficients in the free energy.}
  \label{fig:quadratic_coefficients}
\end{figure}

The diagrams for the quadratic terms in the free energy are shown in Fig.~\ref{fig:quadratic_coefficients}, 
and the leading-order behavior in the limiting cases of vanishing and strong disorder is 
\begin{align}
  a_\mathrm{m}(\vect{q},\omega_m) &= a_\mathrm{m} +\left\{\begin{matrix}\frac{7\zeta(3)\dosf v_\mathrm{F}^2}{8\pi^2T^2}q^2&, \quad T\tau_1\gg 1 \\ 4\dosf v_\mathrm{F}^2 \tau_1^2 q^2&,\quad T\tau_1\ll 1\end{matrix}\right. \nonumber \\ 
  & \qquad + \left\{\begin{matrix}\frac{\pi\dosf}{2T}|\omega_m| &, \quad T\tau_1\gg 1 \\ 8\dosf\tau_1|\omega_m|&,\quad T\tau_1\ll 1\end{matrix}\right. \mathcomma\\ 
  a_{\mathrm{s},11}(\vect{q},\omega_m) &=a_{\mathrm{s},11} + \left\{\begin{matrix} \frac{7\zeta(3)\dosf v_\mathrm{F}^2}{16\pi^2T^2}q^2 &,\quad T\tau_1\gg 1 \\ \frac{\pi\dosf v_\mathrm{F}^2 \tau_1}{8T}q^2 &,\quad T\tau_1\ll 1\end{matrix} \right. \nonumber \\ 
  & \qquad + \frac{\pi\dosf}{4T}|\omega_m| \mathcomma\\
  a_{\mathrm{s},22}(\vect{q},\omega_m) &= a_{\mathrm{s},22} + \frac{7\zeta(3)\dosf v_\mathrm{F}^2}{16\pi^2T^2}q^2+\frac{\pi\dosf}{4T}|\omega_m|\mathcomma \\ 
  a_{\mathrm{s},12}\left(\vect{q},\omega_m\right)  & = a_{\mathrm{s},12} = a_{\mathrm{s},21}\left(\vect{q},\omega_m\right)\mathcomma
\end{align}
where 
\begin{align}
  a_\mathrm{m} &= \tfrac{4}{V_\mathrm{m}} -4\dosf \left[\operatorname{\psi_0}\left(\tfrac{3}{2}+\tfrac{1}{8\pi T\tau_1}+\tfrac{\Lambda_\mathrm{m}}{T}\right)\right. \label{eq:a_m}\nonumber\\ 
  &\qquad -\left.\operatorname{\psi_0}\left(\tfrac{1}{2}+\tfrac{1}{8\pi T\tau_1}\right)\right] \nonumber \\ 
  & \approx \tfrac{4}{V_\mathrm{m}}-4\dosf\ln\tfrac{\Lambda_\mathrm{m}}{\max\left(T, {\tau_1}^{-1}\right)} \mathcomma\\ 
  a_{\mathrm{s},11} &= a_{\mathrm{s},22} = -2\dosf\left[\operatorname{\psi_0}\left(\tfrac{3}{2}+\tfrac{\Lambda_\mathrm{s}}{T}\right)-\operatorname{\psi_0}\left(\tfrac{1}{2}\right)\right] \nonumber \\ 
  &\approx -2\dosf\ln\tfrac{\Lambda_\mathrm{s}}{T} \mathcomma \\ 
  a_{\mathrm{s},12} &= a_{\mathrm{s},21} = -\tfrac{2}{V_\mathrm{s}}\mathperiod
\end{align}
The magnetic critical points $(x,T)$ are the points of the phase diagram (as sketched in Fig.~\ref{fig:phasediagrams}) where $a_\mathrm{m}=0$ holds, and analogously, the superconducting critical points are defined by $a_{\mathrm{s},11}+a_{\mathrm{s},22}+|a_{\mathrm{s},12}|+|a_{\mathrm{s},21}|=0$. Note that doping can affect the values of the coupling constants $V_\mathrm{m}$ and $V_\mathrm{s}$. The intersection of the two critical lines then defines the multicritical point of the phase diagram. 

The renormalization of the superconductive vertex describes diffusion of Cooper pairs in band~1, which, however, does not affect the SC transition temperature because the coefficient $a_{\mathrm{s},1}\equiv a_{\mathrm{s},1}(\vect{0},0)$ does not depend on the scattering rate~${\tau_1}^{-1}$ anymore, and it coincides with the respective result in band~2 which is not affected by impurity scattering. 
For nonmagnetic impurities in a usual $s$-wave superconductor, this constitutes the Anderson theorem\cite{JPhysChemSolids.11.26,JETP.8.1090,JETP.9.220}. 
From Eq.~\eqref{eq:a_m} follows that the magnetic ordering temperature vanishes for ${\tau_1}^{-1}\gtrsim T_\mathrm{N,clean}$, where $T_\mathrm{N,clean}$ is the corresponding transition temperature of the clean system. The phase competition discussed in this paper is, of course, only sensible for ${\tau_1}^{-1}<T_\mathrm{N,clean}$, where both competing states order. 
\begin{table}
\begin{tabular}{|c|c|c|}\hline
  $\xi_\mathrm{SC,AFM}$ & clean & disordered \\ \hline 
  SC & $\frac{v_\mathrm{F}}{T_\mathrm{c}}\left(\frac{T}{T_\mathrm{c}}-1\right)^{-1/2}$ & 
 $\sqrt{\frac{v_\mathrm{F} l}{T_\mathrm{c}}}\left(\frac{T}{T_\mathrm{c}}-1\right)^{-1/2}$  \\ 
  AFM & $\frac{v_\mathrm{F}}{T_\mathrm{c}}\left(\frac{T}{T_\mathrm{c}}-1\right)^{-1/2}$ & $l\left(\frac{T}{T_\mathrm{c}}-1\right)^{-1/2}$ \\ \hline
  \end{tabular}
  \caption{
  The characteristic length scales of the order parameter fluctuations,
  obtained from the Ginzburg-Landau expansion under the assumption of second-order phase transitions at the
  respective critical points.}
  \label{table:length-scales}
\end{table}
From the quadratic coefficients $a_\mathrm{m}$ and $a_{\mathrm{s},11}$ at finite $\vect{q}$ and $\omega_m$ we also find the typical length scales of fluctuations of the magnetic and superconducting order parameters to be affected differently by disorder.  
They are summarized in table~\ref{table:length-scales}. 
In the clean case, we find the same characteristic length scale for both order parameters. 
It corresponds to the result for the coherence length of a superconductor\cite{LarkinVarlamov}, and is independent of disorder strength. 
For strong disorder, both lengths are reduced with increasing scattering rate ${\tau_1}^{-1}$. 
However, the magnetic length is stronger suppressed by disorder than the superconductive coherence
length, cf.~Table~\ref{table:length-scales}. 

Thus, even if the spatial variation of both order parameters is the same in the clean limit, it is different if one includes disorder. Then, the characteristic length scales for the magnetic degrees of freedom become shorter. 
\subsection{Quartic coefficients}
\begin{figure}
   \includegraphics[width=\linewidth]{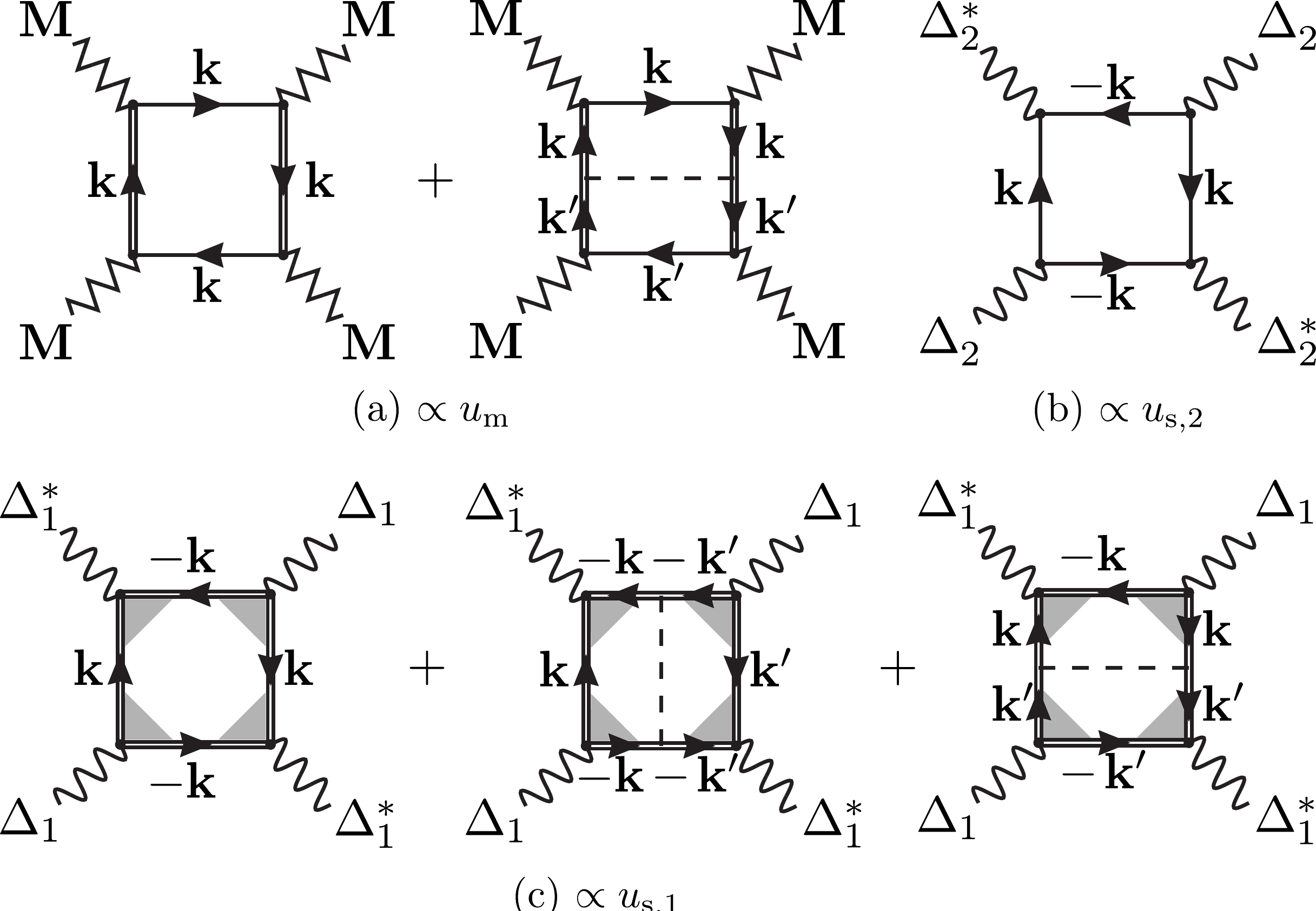}
  \caption{Quartic coefficients I. These contributions correspond to the quartic order terms of pure SC and AFM.}
  \label{fig:quartic-diagrams1}
\end{figure}

The coefficients of the quartic terms of pure SC and AFM, which are not due to phase competition, are depicted in Fig.~\ref{fig:quartic-diagrams1}. 
The resulting coefficients are 
\begin{align}
  u_{\mathrm{m}}&= -\tfrac{\dosf}{4\pi^2 T^2}\operatorname{\psi_2}\left(\tfrac{1}{2}+\tfrac{1}{8\pi T\tau_1}\right) \nonumber \\ 
  & \qquad -\tfrac{\dosf}{96\pi^3 T^3\tau_1}\operatorname{\psi_3}\left(\tfrac{1}{2}+\tfrac{1}{8\pi T\tau_1}\right) \nonumber \\ 
  &= \left\{\begin{matrix} \frac{7\dosf\zeta(3)}{2\pi^2T^2}
 \mathcomma & T\tau_1 \gg 1  \\ \frac{16}{3}\dosf \tau_1^2 \mathcomma & T\tau_1 \ll 1 \end{matrix}\right. \mathcomma \\ 
  u_{\mathrm{s},1} &= u_{\mathrm{s},2} = \frac{7\zeta(3)\dosf}{4\pi^2 T^2} \mathperiod
\end{align}
Again, the coefficients associated with superconductivity do not depend on the strength of disorder whereas the quartic coefficient associated with the magnetic order parameter does. 

\begin{figure}
   \includegraphics[width=0.7\linewidth]{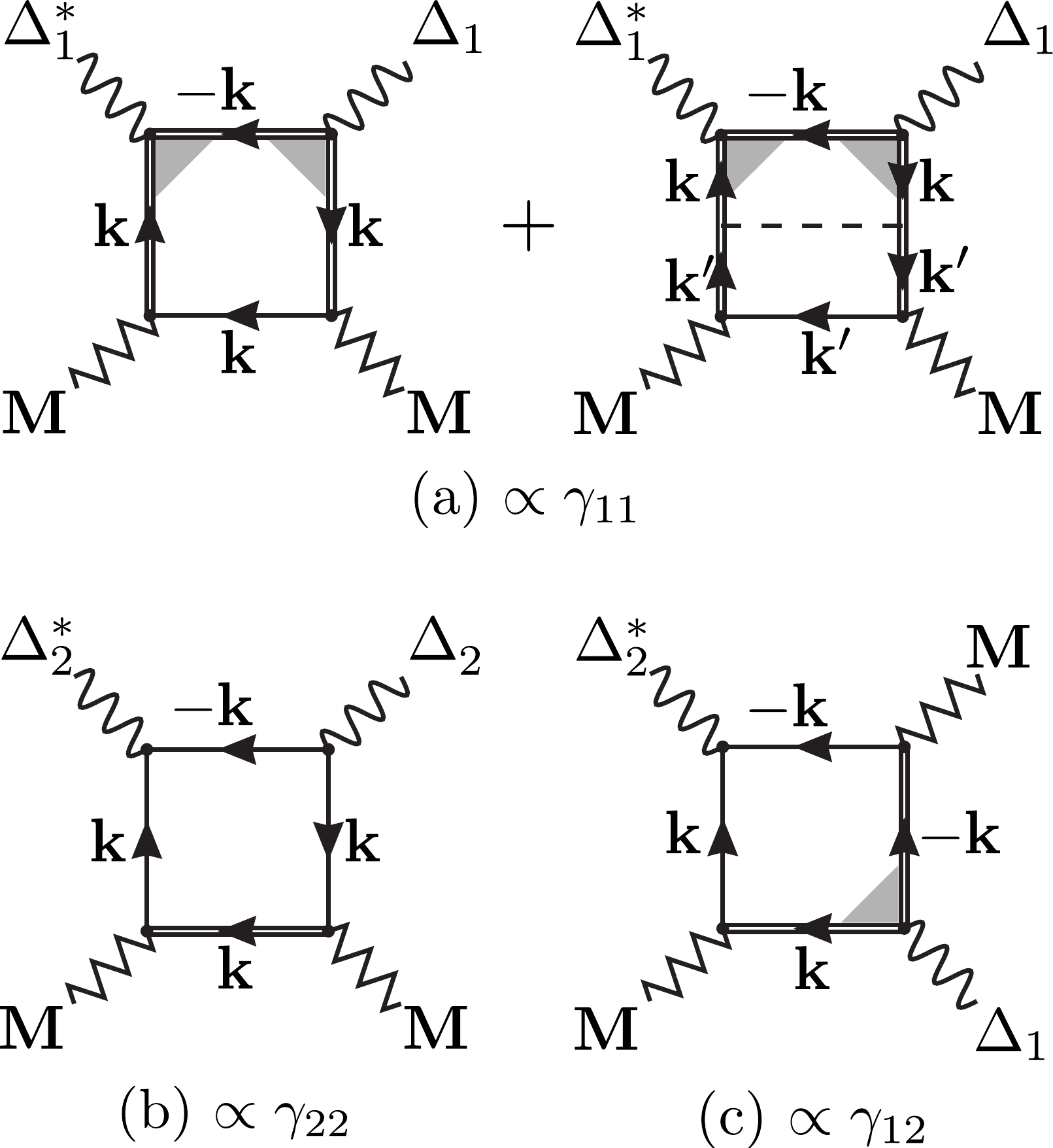}
  \caption{Quartic coefficients II. These contributions are responsible for the phase competition.}
  \label{fig:quartic-diagrams2}
\end{figure}

The diagrams contributing to the coefficients of the quartic terms reflecting the phase competition between magnetic order and superconductivity are depicted in Fig.~\ref{fig:quartic-diagrams2},  
and the results are 
\begin{align}
  \gamma_{11} &= \gamma_{22} = -\tfrac{2\dosf\tau_1}{\pi T}\left[ \operatorname{\psi_1}\left(\tfrac{1}{2}+\tfrac{1}{8\pi T\tau_1}\right)-\operatorname{\psi_1}\left(\tfrac{1}{2}\right) \right]\nonumber \\  
&= \left\{\begin{matrix} 
\frac{7\zeta(3)\dosf}{2\pi^2T^2} \mathcomma & T\tau_1\gg 1 \\ \frac{\pi\dosf\tau_1}{T} \mathcomma & T\tau_1\ll 1  \end{matrix}\right. \mathcomma \\ 
  \gamma_{12} &= \gamma_{21} = -16\dosf\tau_1^2 \left[\operatorname{\psi_0}\left(\tfrac{1}{2}+\tfrac{1}{8\pi T\tau_1}\right) -\operatorname{\psi_0}\left(\tfrac{1}{2}\right)\right] \nonumber \\ 
  &\qquad + \tfrac{2\dosf\tau_1}{\pi T}\operatorname{\psi_1}\left(\tfrac{1}{2}\right) \nonumber \\
  &= \left\{ \begin{matrix} \frac{7\zeta(3)\dosf}{4\pi^2T^2} \mathcomma &  T\tau_1 \gg 1 \\ \frac{\pi\dosf\tau_1}{T} \mathcomma & T\tau_1\ll 1   \end{matrix} \right. \mathperiod
\end{align}
Depending on the assumption for the underlying symmetry of the superconducting order parameter, these results can be summarized into 
\begin{align}
  \gamma_{++}&=\gamma_{11}+\gamma_{22}+\gamma_{12}+\gamma_{21} \nonumber \\ 
  &= \left\{ \begin{matrix} 3\frac{7\zeta(3)\dosf}{2\pi^2T^2}
\mathcomma & T\tau_1\gg 1 \\ \frac{4\pi\dosf\tau_1}{T} \mathcomma & T\tau_1\ll 1 \end{matrix} \right. \mathcomma \\ 
  \gamma_{+-}&=\gamma_{11}+\gamma_{22}-\gamma_{12}-\gamma_{21}\nonumber \\ 
  &= \left\{ \begin{matrix} \frac{7\zeta(3)\dosf}{2\pi^2T^2}
\mathcomma & T\tau_1\gg 1 \\ 32\dosf\tau_1^2\mathcomma & T\tau_1\ll 1 \end{matrix} \right. \mathcomma
  \end{align}
where the indices refer to $s^{++}$ and $s^{+-}$ symmetry of the order parameter $\Delta$, and we omitted the expressions for arbitrary $T\tau_1$ for the sake of brevity. 

In the case of strong disorder, we find $\gamma_{+-}\rightarrow 0$ to leading order, i.\,e., superconducting and magnetic order parameters completely 
decouple in the limit ${\tau_1}^{-1}\rightarrow\infty$, thus competition between magnetism and superconductivity ceases to exist in this limit. 
As the consideration of finite interband scattering shows, this complete decoupling only occurs in the limit ${\tau_{12}}^{-1}\rightarrow 0$, 
but even at finite interband scattering rates, the competition between SC and AFM order is mitigated by the intraband scattering. 
Since we are in the regime of weak disorder, the limit of large ${\tau}^{-1}$ is understood in the sense that $T_\mathrm{c}\ll\tau^{-1}\ll E_\mathrm{F}$, where $E_\mathrm{F}$ denotes the Fermi energy. 

In the case of zero interband scattering rate, these calculations can be easily generalized to arbitrary finite intraband scattering rates ${\tau_1}^{-1}$ and ${\tau_2}^{-1}$ in band~1 and~2, respectively. 
Our calculations show that the intraband scattering rates in the absence of interband scattering simply add up to a total scattering rate ${\tau_t}^{-1}={\tau_1}^{-1}+{\tau_2}^{-1}$. 
Therefore, the structure of the resulting coefficients remains the same, and the corresponding coefficients can be obtained by substitution of ${\tau_t}^{-1}$ for ${\tau_1}^{-1}$ in the previously discussed expansion of the free energy. 
\section{Phase competition in presence of intraband scattering}\label{sec:intraband}
\begin{figure}[t]
    \includegraphics[width=0.23\textwidth]{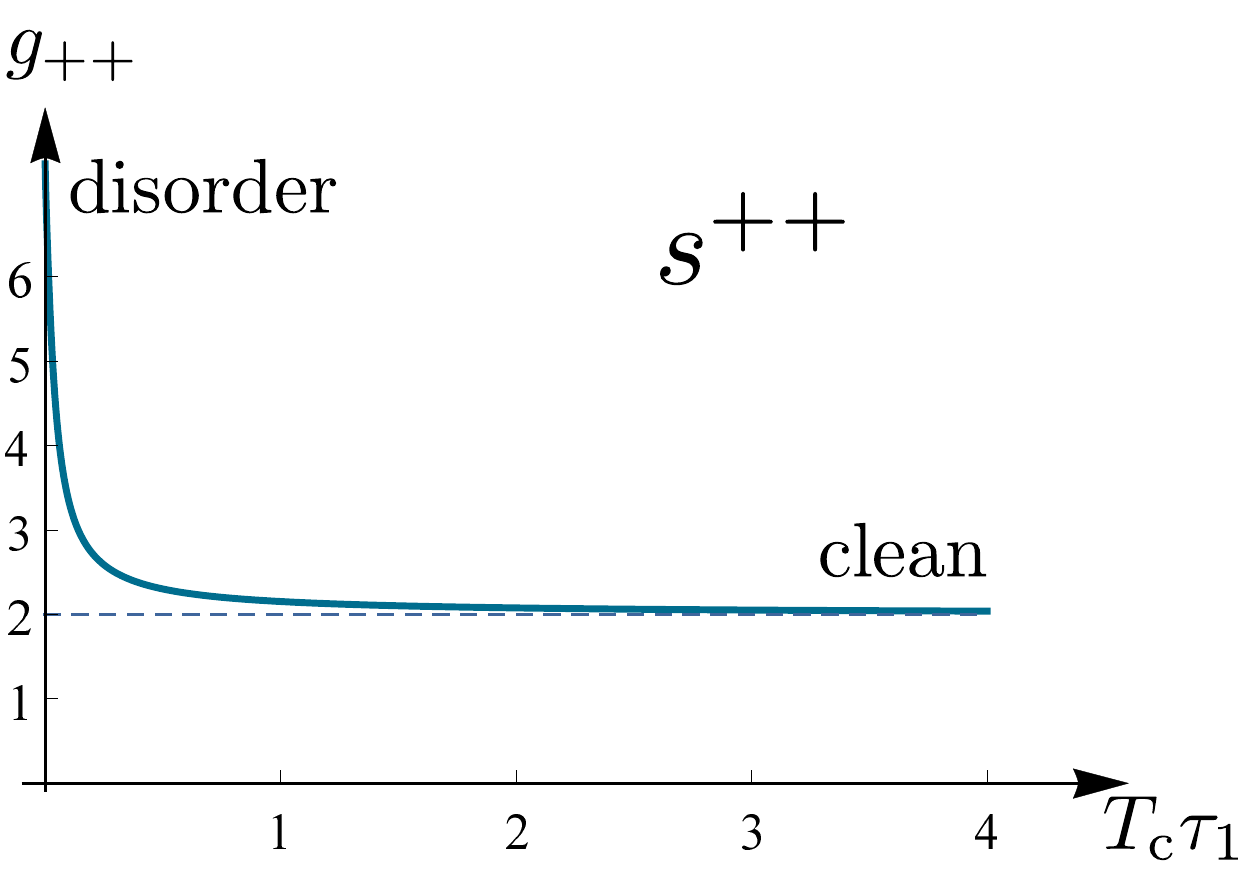}
    \includegraphics[width=0.23\textwidth]{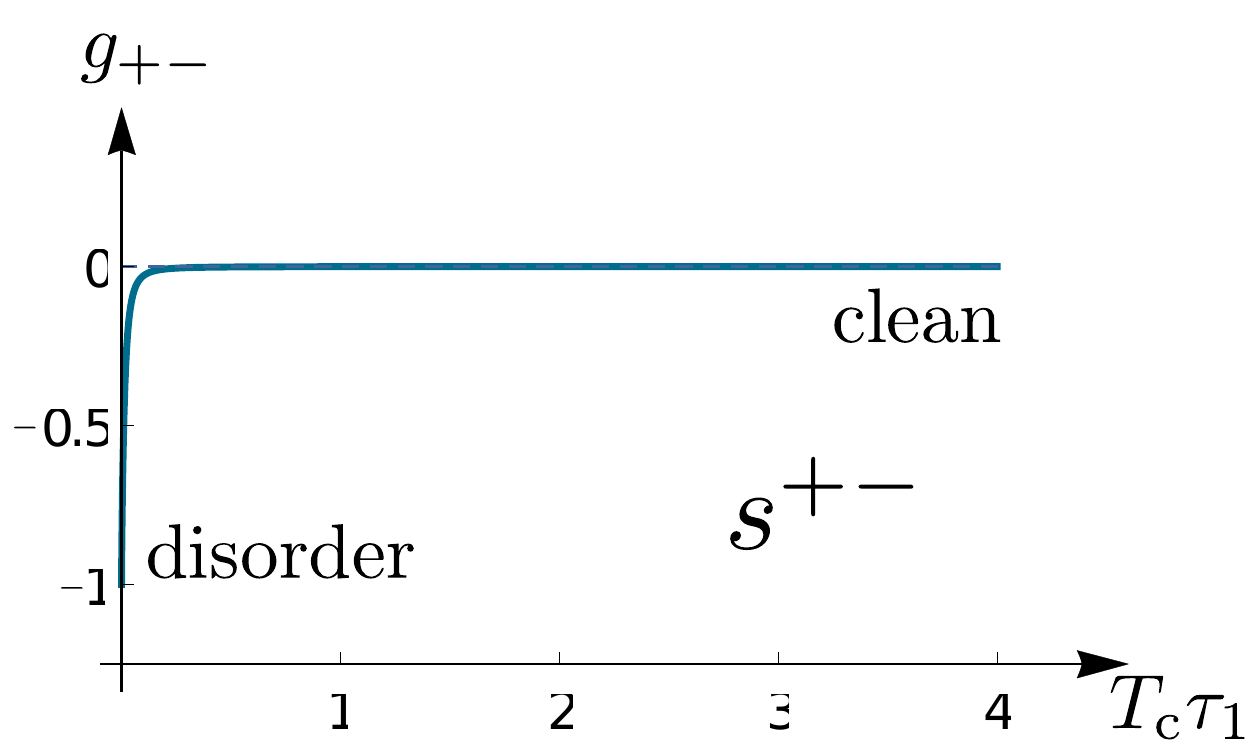}
  \caption{$g$ as a function of $T_\mathrm{c}\tau$ for $s^{++}$ and $s^{+-}$ pairing symmetries.}
  \label{fig:g}
\end{figure}

The full Ginzburg-Landau expansion now allows us to calculate $g_{+\pm}=\gamma_{+\pm}/\sqrt{u_\mathrm{s}u_\mathrm{m}}-1$ which determines the nature of the phase diagram. Here, the index refers to the respective symmetry of the SC~order parameter. 
In Fig.~\ref{fig:g}, we plotted $g$ as a function of $T_\mathrm{c}\tau_1$ for the two pairing symmetries under consideration, since we expanded the free energy around the multicritical point, where $T\approx T_\mathrm{c}$. 
In the two limiting cases of vanishing disorder and strong disorder, we find
\begin{align}
  g_{++}&=\frac{\gamma_{++}}{\sqrt{u_\mathrm{m}u_\mathrm{s}}}-1 \nonumber\\ 
  &=\left\{ \begin{matrix} 2 \mathcomma &T_\mathrm{c}\tau_1 \gg 1 \\ \pi^2\sqrt{\frac{6\zeta(3)}{7}}-1 \approx 7.3 \mathcomma & T_\mathrm{c}\tau_1\ll 1 \end{matrix} \right. \mathcomma \\
  g_{+-}&=\frac{\gamma_{+-}}{\sqrt{u_\mathrm{m}u_\mathrm{s}}}-1 \nonumber\\
  &= \left\{ \begin{matrix} 0 \mathcomma & T_\mathrm{c}\tau_1 \gg 1  \\ -1 \mathcomma & T_\mathrm{c}\tau_1 \ll 1 \end{matrix} \right. 
\end{align}
for the two respective cases. 
In the clean limit ($T_\mathrm{c}\tau_1\rightarrow \infty$) we recover the results obtained from a model disregarding disorder\cite{PhysRevB.81.140501,PhysRevB.81.174538,PhysRevB.82.014521}. $g_{++}=2$ means that $s^{++}$~superconductivity cannot coexist with antiferromagnetism, whereas $g_{+-}=0$ allows for both possible types of phase diagrams 
since a more detailed band structure may lead to a small positive or negative $g_{+-}$. This was taken as evidence against $s^{++}$~superconductivity to be realized in the iron pnictides.

\begin{figure}[t]
  \includegraphics[width=0.4\textwidth]{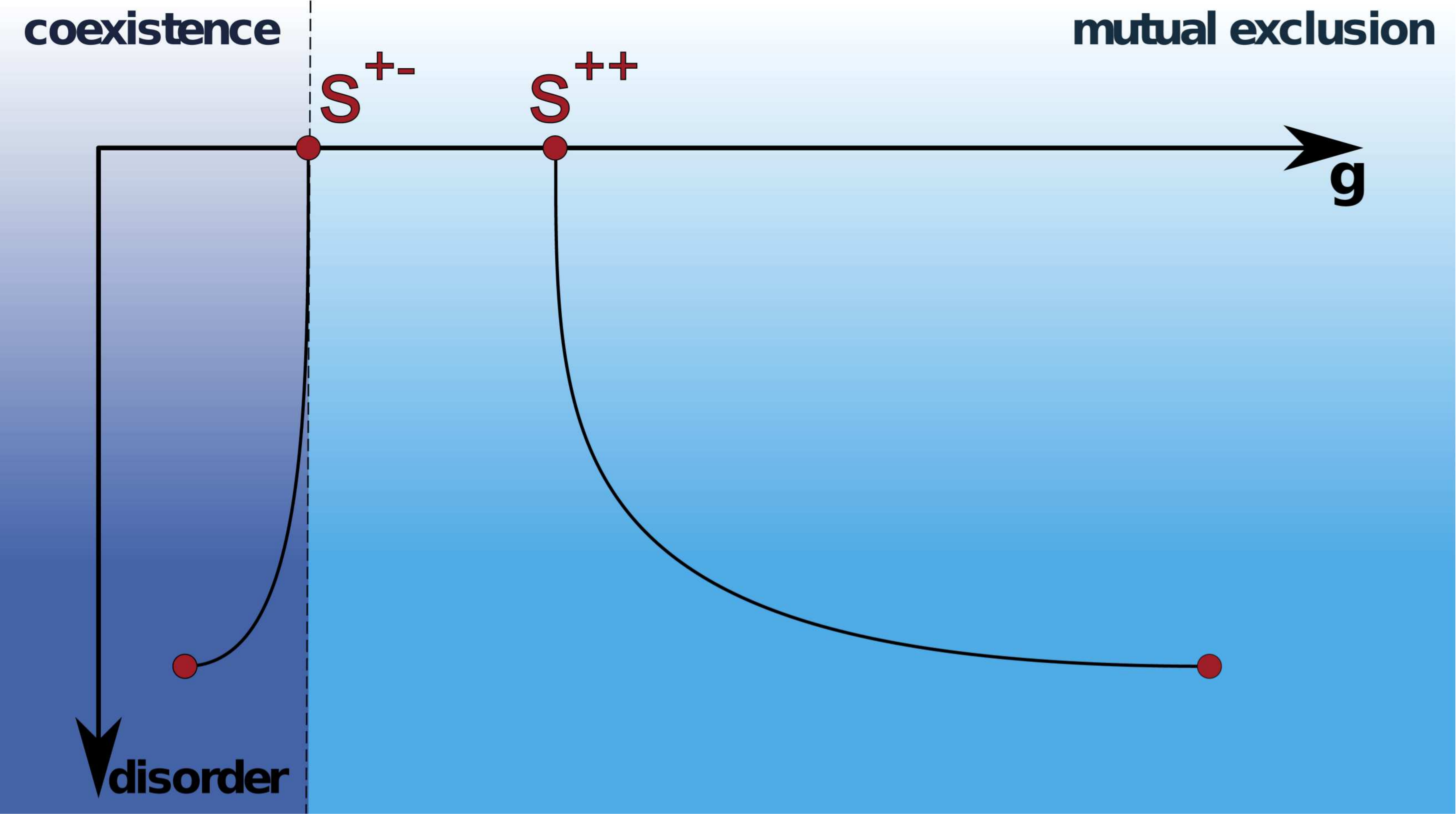}
  \caption{(Color online) Influence of disorder on $g$ in case of $s^{+-}$ and $s^{++}$ pairing.}
  \label{fig:result}
\end{figure}

The consideration of disorder supports this reasoning since $g_{++}$ increases with disorder and the $s^{++}$ pairing state is even more inconsistent with the observed variety of phase diagrams. The $s^{+-}$~pairing state is driven towards the regime of phase coexistence by increasing disorder 
but the consideration of a more detailed band structure would still allow for both types of phase diagrams. 
The result of $g_{+-}=-1$ in the limit of strong disorder results from the complete decoupling of AFM and SC orders since in this limit $\gamma_{+-}=0$. 
Note that our findings also imply the possibility of a disorder-induced transition from mutual exclusion to coexistence of superconductivity and antiferromagnetism in the iron pnictides. 
These findings are summarized in Fig.~\ref{fig:result}. 
\section{Influence of interband scattering}\label{sec:interband}
Our treatment of the problem with a finite interband scattering rate ${\tau_{12}}^{-1}$ in Eilenberger formalism\cite{Eilenberger1968} is analogous to the approach described in Ref.~\onlinecite{PhysRevB.84.214521}, 
and we consider a model with finite interband scattering rate ${\tau_{12}}^{-1}$ and equal intraband scattering rates ${\tau_0}^{-1}$ in the hole and electron bands as well. 
This approach allows for the expansion of the SC and AFM gap equations which can, up to a factor, be identified with the first derivative of the free energy with respect to the order parameters. 

We extract the coefficients of the free energy expansion from the expanded gap equations. 
In presence of interband scattering, the coefficients contributing to the quantity $g_{+\pm}=\gamma_{+\pm}/\sqrt{u_{\mathrm{s},+\pm}u_\mathrm{m}}-1$ which determines the nature of the multicritical point, read 
\begin{widetext}
  \begin{align}
  u_{\mathrm{s},+-}&= -\tfrac{\dosf}{4\pi^2 T^2}\operatorname{\psi_2}\left(\tfrac{1}{2}+\tfrac{t}{4\pi T\tau_0}\right)-\tfrac{\dosf}{12\pi^2 T^2}\tfrac{t}{4\pi T\tau_0}\operatorname{\psi_3}\left(\tfrac{1}{2}+\tfrac{t}{4\pi T\tau_0}\right) \label{eq:interband-us+-} 
  \mathcomma \\ 
  u_{\mathrm{s},++}&=-\tfrac{\dosf}{4\pi^2 T^2}\operatorname{\psi_2}\left(\tfrac{1}{2}\right)=\tfrac{7\dosf\operatorname{\zeta}(3)}{2\pi^2 T^2}\mathcomma \\
  u_\mathrm{m}&= -\tfrac{\dosf}{4\pi^2 T^2}\operatorname{\psi_2}\left(\tfrac{1}{2}+\tfrac{1+t}{4\pi T\tau_0}\right)-\tfrac{\dosf}{12\pi^2 T^2}\tfrac{1+t}{4\pi T\tau_0}\operatorname{\psi_3}\left(\tfrac{1}{2}+\tfrac{1+t}{4\pi T\tau_0}\right) 
  \mathcomma\\
    \gamma_{+-}&=8\dosf\tau_0^2\left(1+2t\right)\left[\operatorname{\psi_0}\left(\tfrac{1}{2}+\tfrac{1+t}{4\pi T\tau_0}\right)-\operatorname{\psi_0}\left(\tfrac{1}{2}+\tfrac{t}{4\pi T\tau_{0}}\right)\right] \nonumber \\ 
    &\qquad -\tfrac{2\dosf\tau_0}{\pi T}\left[\left(1+t\right)\operatorname{\psi_1}\left(\tfrac{1}{2}+\tfrac{1+t}{4\pi T\tau_0}\right)+t\operatorname{\psi_1}\left(\tfrac{1}{2}+\tfrac{t}{4\pi T\tau_{0}}\right)\right]
    \mathcomma \\
    \gamma_{++}&= -\tfrac{8\dosf\tau_0^2}{(1+t)^2}\left[\operatorname{\psi_0}\left(\tfrac{1}{2}+\tfrac{1+t}{4\pi T\tau_0}\right)-\operatorname{\psi_0}\left(\tfrac{1}{2}\right)\right] 
    -\tfrac{2\dosf\tau_0}{\pi T}\tfrac{1}{1+t}\left[\operatorname{\psi_1}\left(\tfrac{1}{2}+\tfrac{1+t}{4\pi T\tau_0}\right)-2\operatorname{\psi_1}\left(\tfrac{1}{2}\right)\right]
    \mathcomma 
    \label{eq:interband-gamma++}
  \end{align}
\end{widetext}
where $t=\tau_{0}/\tau_{12}<\infty$ is the ratio of interband to intraband scattering rate. 
In the limit of ${\tau_{12}}^{-1}=0$, this corresponds to the results discussed in section~\ref{sec:intraband}, 
but with the intraband scattering rate doubled, since now the scattering rate in the electron band is finite, and equal to the scattering rate in the hole band.  Again, we find that AFM and $s^{+-}$~SC decouple in the limit of $T_\mathrm{c}\tau_0\rightarrow\infty$ ($g_{+-}=-1$), resulting in a phase diagram exhibiting a region where AFM and SC coexist microscopically whereas for $s^{++}$~SC, $g_{++}\approx 7.3$, and AFM and SC exclude each other. 
\begin{figure*}
  \includegraphics[width=0.8\linewidth]{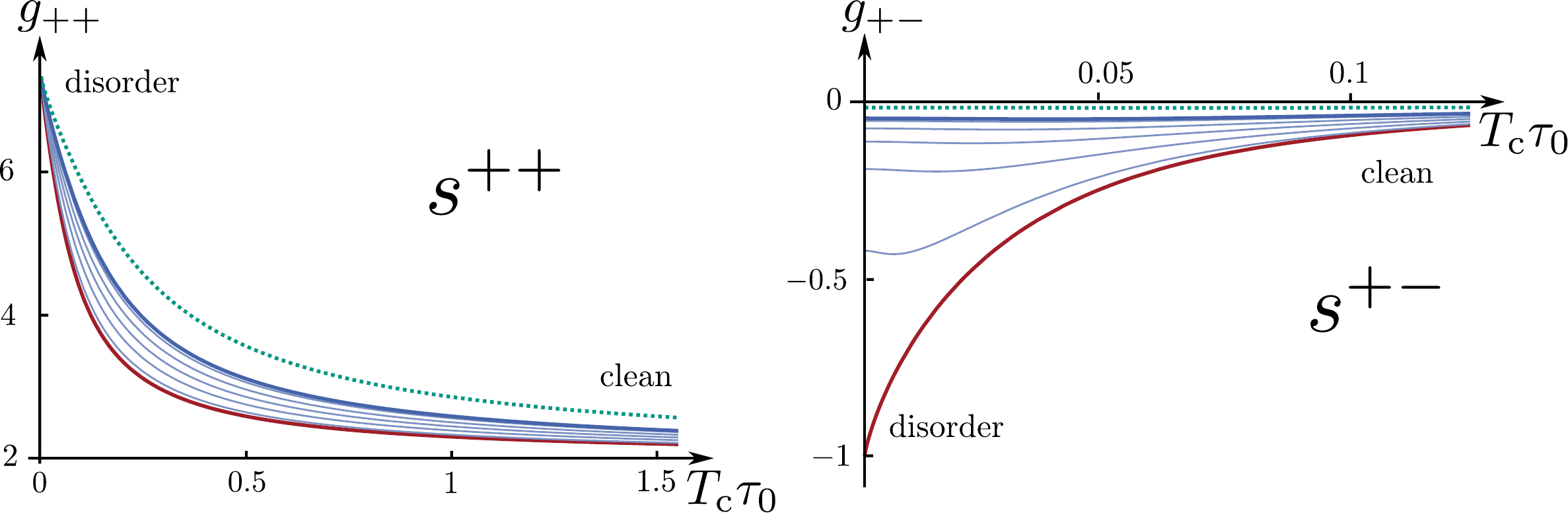}
  \caption{(Color online) The parameter $g$ which characterizes the shape of the phase diagram as a function of $T_\mathrm{c}\tau_0$ for different ratios of interband to intraband scattering rate $t$: $t=0$ (red line), $0<t<1$ (light blue lines), $t=1$ (blue line), and $t=2$ (green dotted line).}
  \label{fig:result-interband}
\end{figure*}

For finite interband scattering rates, AFM and SC no longer decouple completely in the limit of strong disorder but the coupling $\gamma_{+-}$ is reduced in the case of $s^{+-}$~pairing compared to the clean case. The stronger interband scattering is, the closer $g_{+-}$ is to zero, but for all ratios $t>0$ we found $g_{+-}<0$, implying a phase diagram with a tetracritical point and coexistence of SC and AFM. 
For $s^{++}$~pairing, in contrast, we find $\lim_{T_\mathrm{c}\tau_0\rightarrow 0}g_{++}\approx7.3$ and $\lim_{T_\mathrm{c}\tau_0\rightarrow0}=2$, irrespective of the ratio~$t$. 
Furthermore, we found that in case of $s^{++}$~pairing, the interband scattering simply adds to the intraband scattering rate. 
Thus, the qualitative behavior does not depend on $t$, and we find $g\geq 2$, thus $s^{++}$~SC is not able to coexist microscopically with AFM, not even in the presence of inter- and/or intraband scattering. These results are summarized in Fig.~\ref{fig:result-interband}. 

In conclusion, for the $s^{+-}$~pairing state, intraband scattering and interband scattering are antagonistic processes, but the effect of intraband scattering is always stronger, even in the limit of $t>1$, whereas for the $s^{++}$~pairing state, the rates of these two scattering processes simply add up. 
Thus in the presence of interband scattering, we find the $s^{++}$ pairing state to be inconsistent with phase diagrams revealing a regime of microscopic coexistence of AFM and SC. The analysis of the $s^{+-}$~state, in contrast, suggests that this order parameter symmetry will always result in a phase diagram exhibiting a regime of coexistence of AFM and SC. Still, a more detailed analysis including finite ellipticity and/or chemical potential could also lead to $g>0$, thus allowing for both types of phase diagrams. Since ellipticity and chemical potential yield only small corrections, they could not bring the $s^{++}$~state to coexists with magnetic order. 
Therefore, the analysis of finite inter- and intraband scattering supports the reasoning based on phase competition against the $s^{++}$ as a suitable candidate for the pairing state in iron pnictides. 
\section{Conclusion} \label{sec:conclusions}
We studied a model of iron pnictides and related iron-based superconductors and included impurity scattering in the microscopic model. We developed two complementary simplified models for impurity scattering in the iron pnictides motivated by experimental observations. 
Model~A concentrates on the most important scattering process in the materials under consideration which is intraband scattering in the hole band. Therefore, we neglected intraband scattering in the electron band and all types of interband scattering processes, since ${\tau_1}^{-1}\gg{\tau_2}^{-1},{\tau_{12}}^{-1}$. 
Model~B focuses on the interband scattering rate, neglected in model~A, yet makes the simplifying assumption that the intraband scattering rates of both bands are the same. 

We derived the full Ginzburg-Landau expansion of the free energy from this microscopic model. 
From the quadratic coefficients we find that the transition temperature of neither $s^{++}$ nor $s^{+-}$~superconductivity is influenced by impurity scattering if we take only intraband scattering into account. This is in accordance with the Anderson theorem. 
We further compared the coherence lengths obtained from the gradient terms in the expansion and found the coherence length of the magnetic order parameter more strongly reduced by impurity scattering than the coherence length of the superconducting order parameter. 

From the quartic coefficients we concluded how impurity scattering affects the phase competition in the iron pnictides. Our analysis supports the argument obtained in the clean case\cite{PhysRevB.81.140501,PhysRevB.81.174538,PhysRevB.82.014521} that $s^{++}$~superconductivity is inconsistent with phase diagrams that show microscopic coexistence of antiferromagnetism and superconductivity. 
This behaviour occurs in models with and without interband scattering. 
Thus the consideration of disorder provides an even stronger argument against the $s^{++}$ state to be realized in the iron pnictides and supports  $s^{+-}$~superconductivity or other sign-changing superconducting states. 
\section*{Acknowledgments}
We thank A.~V.~Chubukov, R.~M.~Fernandes, and P.~Hirschfeld for helpful discussions. This work was supported by the Deutsche Forschungsgemeinschaft through DFG-SPP~1458 `Hochtemperatursupraleitung in Eisenpniktiden'.
The work of SVS was partially supported by the Alexander von Humboldt Foundation through Feodor Lynen Research
Fellowship and by the NSF grants DMR-1001240 and PHY-1125844.
\appendix
\section{Exemplary calculation of $\gamma_{12}$}\label{app:examplecoefficient}
To illustrate the calculation of the diagrams for the coefficients in the free energy, let us provide here
a detailed computation of the coefficient $\gamma_{12}$ in the framework of our simplified model. 
In the absence of interband scattering there is one diagram that contributes to the coefficient $\gamma_{12}$. In the following, we use the abbreviation $\tilde{\nu}_n=\nu_n+\sgn\nu_n /2\tau_1$ and use that the Cooperon ladder only depends on the absolute value of $\nu_n$. 
The diagram, Fig.~6c, evaluates to
\begin{align} 
  & T\sum_{n=-\infty}^\infty\int\frac{\mathrm{d}\vect{k}}{\left(2\pi\right)^2}\,C_1(\nu_n) \nonumber \\ 
  &\qquad \times G_{1,\vect{k}}(\nu_n)G_{1,-\vect{k}}(-\nu_n)G_{2,\vect{k}}(\nu_n)G_{2,-\vect{k}}(-\nu_n) \nonumber \\ 
  &=T\sum_{n=-\infty}^\infty C_1(\nu_n)\dosf\int\mathrm{d}\epsilon\,\frac{1}{\epsilon-\mathrm{i}\tilde{\nu}_n}\frac{1}{\epsilon+\mathrm{i}\tilde{\nu}_n}\frac{1}{\epsilon+\mathrm{i}\nu_n}\frac{1}{\epsilon-\mathrm{i}\nu_n}\nonumber \\ 
  &= \dosf T\sum_{n=-\infty}^\infty \frac{4\pi\tau_1^2C_1(\nu_n)}{|\nu_n|\left(2\tau_1|\nu_n|+1\right)\left(4\tau_1|\nu_n|+1\right)} \nonumber \\
  &= \frac{\dosf}{4\pi^2T^2}\sum_{n=0}^\infty\frac{1}{\left(n+\frac{1}{2}\right)^2\left(n+\frac{1}{2}+\frac{1}{8\pi T\tau_1}\right)} \mathperiod
\end{align} 
This sum may be conveniently evaluated approximately in the limits $T\tau_1\gg 1$ and $T\tau_1\ll 1$,
as well as exactly. To calculate the coefficients from the diagrams we have to include the proper symmetry factor which is 2 in the case of $\gamma_{12}$.
The resulting coefficient is then given by
\begin{align}
  \gamma_{12}&=\frac{\pi\dosf\tau_1}{T}+16\dosf{\tau_1}^2\left[\operatorname{\psi_0}\left(\tfrac{1}{2}\right)-\operatorname{\psi_0}\left(\tfrac{1}{2}+\tfrac{1}{8\pi T\tau_1}\right)\right] \nonumber \\ 
    &= \left\{\begin{matrix} \frac{7\operatorname{\zeta}(3)\dosf}{4\pi^2 T^2} &\mathcomma \quad T\tau_1\gg 1\mathcomma \\ \frac{\pi\dosf\tau_1}{T}  &\mathcomma\quad T\tau_1\ll 1\mathperiod \end{matrix} \right.
\end{align}
\section{Treatment of AFM and $s^{++}$~SC in the Eilenberger formalism}\label{app:eilenberger-s++}
The coefficients of the Ginzburg-Landau expansion for model~B, given in Eq.~\eqref{eq:interband-us+-} to~\eqref{eq:interband-gamma++}, have been obtained from the equation of state using the Eilenberger formalism~\cite{Eilenberger1968}. 
The application of the Eilenberger formalism to a system showing antiferromagnetism and $s^{+-}$~superconductivity can be done in complete analogy to Ref.~\onlinecite{PhysRevB.84.214521}, using the same parametrization of the Eilenberger Green's function. In this appendix, we sketch the respective procedure for $s^{++}$~superconductivity. The mean-field Hamiltonian can be written as $\mathcal{H}=\tfrac{1}{2}\sum_{\vect{k},\alpha,\beta}\bar{\Psi}_{\vect{k},\alpha}H_{\vect{k},\alpha\beta}\Psi_{\vect{k},\beta}$ where we summarized the fermionic operators in the two bands into $\bar{\Psi}_{\vect{k},\alpha}=(\begin{matrix}\psi_{1,\vect{k},\alpha}^\dagger&\psi_{1,-\vect{k},\alpha}&\psi_{2,\vect{k},\alpha}^\dagger&\psi_{2,-\vect{k},\alpha}\end{matrix})$ and introduced the Hamiltonian matrix consisting of noninteracting and mean-field parts, 
\begin{align}
  H_{\vect{k}}&=H_{0,\vect{k}}+H_{\mathrm{mf},\vect{k}}\nonumber\\ 
  &=\xi_{\vect{k}}\tau_3\rho_3\sigma_0-\Delta\tau_2\rho_0\sigma_2+M\tau_3\rho_1\sigma_3 \mathcomma
  \label{eq:Hamiltonian-s++-Eilenberger}
\end{align}
where $\tau_i$, $\rho_i$, and $\sigma_i$ are the Pauli matrices in Nambu, band, and spin space, respectively. 
The matrix Green's function is defined by 
\begin{equation}
  (\mathrm{i}\nu_n-H_{\vect{k}}-\Sigma)G(\vect{k},\nu_n)=\mathds{1}
  \label{eq:definition-G}
\end{equation}
where the self energy of model~A (intraband scattering with rate ${\tau_0}^{-1}$ in both bands, interband scattering with rate ${\tau_{12}}^{-1}$) is given by 
\begin{align}
  \Sigma&=\frac{1}{4\pi\dosf\tau_0}\int\frac{\mathrm{d}\vect{k}}{(2\pi)^2}\,\tau_3\rho_0\sigma_0G(\vect{k},\nu_n)\tau_3\rho_0\sigma_0 \nonumber\\
  &\qquad+\frac{1}{4\pi\dosf\tau_{12}}\int\frac{\mathrm{d}\vect{k}}{(2\pi)^2}\tau_3\rho_1\sigma_0G(\vect{k},\nu_n)\tau_3\rho_1\sigma_0 \mathperiod
  \label{eq:definition-sigma}
\end{align}
Since in the gap equations as well as in the self energy, the matrix Green's function only appears integrated over momenta, it is convenient to introduce the 
Eilenberger (or quasiclassical) Green's function 
\begin{equation}
  \mathcal{G}(\nu_n)=\frac{2\mathrm{i}}{\pi\dosf}\int\frac{\mathrm{d}\vect{k}}{(2\pi)^2}\,\tau_3\rho_3\sigma_0G(\vect{k},\nu_n)\mathcomma 
  \label{eq:definition-EilenbergerG}
\end{equation}
and rewrite all the equations in terms of $\mathcal{G}$. 
From Eqs.~\eqref{eq:Hamiltonian-s++-Eilenberger} to~\eqref{eq:definition-EilenbergerG}, we find self-consistently that for our model, the Eilenberger Green's function must be of the form 
\begin{equation}
  \mathcal{G}=g_{\nu_n}\tau_3\rho_3\sigma_0-\mathrm{i}f_{\nu_n}\tau_1\rho_3\sigma_2-\mathrm{i}s_{\nu_n}\tau_0\rho_2\sigma_3+o_{\nu_n}\tau_2\rho_2\sigma_1\mathperiod
  \label{eq:parametrization-s++}
\end{equation}
This parametrization can be used to obtain an expansion of the gap equations for the order parameters $\Delta$ and $M$. Such an expansion corresponds, up to an overall prefactor, to the first derivative of the free energy with respect to the order parameter, i.e., the equation of state. Therefore, we can extract the quadratic and quartic coefficients of the free-energy expansion. 

By using the parametrization~\eqref{eq:parametrization-s++} in the self energy~\eqref{eq:definition-sigma}, we find that the self energy depends on intraband and interband scattering rate only via the total scattering rate ${\tau_t}^{-1}\equiv{\tau_0}^{-1}+{\tau_{12}}^{-1}$ for the $s^{++}$~pairing state. Thus we expect no qualitatively new effects due to interband scattering here that were not already captured in the analysis of intraband scattering. 

\end{document}